\documentstyle[12pt]{article}
\begin{document}
\begin{flushright}
hep-th/0311193\\
SNBNCBS-2003
\end{flushright}
\vskip 1.5cm
\begin{center}
{\bf \Large { Superfield Approach to
Symmetries for Matter Fields \\ in Abelian Gauge Theories}}

\vskip 3.5cm

{\bf R.P.Malik}
\footnote{ E-mail address: malik@boson.bose.res.in  }\\
{\it S. N. Bose National Centre for Basic Sciences,} \\
{\it Block-JD, Sector-III, Salt Lake, Calcutta-700 098, India} \\

\vskip 2.5cm

\end{center}

\noindent
{\bf Abstract}:
The derivation of the nilpotent Becchi-Rouet-Stora-Tyutin (BRST)-
and anti-BRST  symmetries for the matter fields,
present in any arbitrary interacting gauge theory, has been a long-standing 
problem in the framework of superfield approach to BRST formalism.
These nilpotent (anti-)BRST symmetries for the Dirac fields are derived
in the superfield formulation for the interacting Abelian gauge theory in 
four $(3 + 1)$-dimensions (4D) of spacetime. The same type of symmetries 
are deduced for the 4D complex scalar fields having a gauge invariant 
interaction with the $U(1)$ gauge field. The above interacting theories are 
considered on a six $(4 + 2)$-dimensional supermanifold
parametrized by four {\it even} spacetime coordinates and a couple of
{\it odd} elements of the Grassmann algebra. The invariance of the 
conserved matter (super)currents and the horizontality condition on the 
(super)manifolds play very important roles in the above derivations. The 
geometrical origin and interpretation for all the above off-shell 
nilpotent symmetries are provided in the framework of superfield formalism.\\

\baselineskip=16pt



\newpage

\noindent
{\bf 1 Introduction}\\

\noindent
The Becchi-Rouet-Stora-Tyutin (BRST) formalism is 
one of the most elegant methods for the covariant canonical quantization
of the gauge theories as well as the reparametrization invariant theories
which are endowed with the first-class constraints in the language of the 
Dirac's prescription for the classification of constraints [1,2]. In the realm
of frontier areas of research connected with  topological field 
theories [3-5] and (super)string theories (see, eg, [6,7] and references
therein), the reach and range of the applicability of
BRST formalism are overwhelming.
The scope of this formalism has been beautifully extended to encompass 
the second-class constraints in its domain of applications [8]. Its
geometrical interpretation in the framework of superfield formulation
and its intimate connections with the basic tenets of supersymmetry [9-14], 
its mathematically
consistent inclusion in the Batalin-Vilkovisky formalism [15,16], its
deep connections with the basic ideas behind
the differential geometry and cohomology [17-21], etc.,
have elevated the subject of BRST formalism to a high degree of 
mathematical sophistication and very useful physical applications. The
true strength of the BRST formalism appears in its full glory in the 
context of interacting non-Abelian gauge theories where the unitarity and
``quantum'' gauge (i.e. BRST) invariance are respected together at 
any arbitrary order of perturbative computation for a given
physical process involving the matter fields and the
non-Abelian gauge fields (which are clearly the true
physical fields of the theory). In this context, it is pertinent to 
point out that 
for each loop diagram consisting of the (gluon) gauge fields,
there exists a corresponding loop diagram consisting of the (anti-)ghost 
fields (which are not the physical fields of the theory in the true sense)
so that the unitarity can be maintained for a given physical process
(see, eg, [22] for details).

In our present endeavour, we shall be concentrating {\it only} on 
the key points associated with the geometrical aspects of the
superfield approach applied
to BRST formalism. This superfield technique is one of the
most interesting and intuitive approaches to gain an insight into the
geometrical meaning of the conserved and nilpotent $(Q_{(a)b}^2 = 0$)
(anti-)BRST charges as well as the nilpotent $(s_{(a)b}^2 = 0$)
(anti-)BRST symmetries they generate for 
the Lagrangian density of a given $p$-form ($ p = 1, 2, 3...)$
gauge theory defined on the $D$-dimensional spacetime manifold. The key idea in
this formulation is to consider the $D$-dimensional $p$-form gauge theory
on a $(D + 2)$-dimensional supermanifold parametrized by the $D$-number
of spacetime (even) coordinates $x^\mu (\mu = 0, 1, 2, 3.....D-1)$ and
a couple of Grassmannian (odd) variables $\theta$ and $\bar \theta$
(with $\theta^2 = \bar\theta^2 = 0, 
\theta \bar\theta + \bar\theta \theta = 0$). One constructs the 
super curvature $(p +1)$-form $\tilde F = \tilde d \tilde A + \tilde A \wedge
\tilde A$ from the super exterior derivative $\tilde d$ (with 
$\tilde d^2 = 0$) and the super connection $p$-form $\tilde A$. This is finally
equated, due to the so-called 
{\it horizontality condition}, with the ordinary 
curvature $(p + 1)$-form $F = d A + A \wedge A$ constructed from the 
ordinary exterior 
derivative $d = dx^\mu \partial_\mu$ (with $d^2 = 0$) and the $p$-form 
ordinary connection $A$.
The above restriction is referred to as the {\it soul-flatness} condition
in [23] which amounts to setting equal to zero all the Grassmannian
components of the $(p + 1)$-rank (anti-)symmetric curvature tensor
that is required in the definition of the $(p + 1)$-form 
super curvature on the $(D + 2)$-dimensional supermanifold. The
procedure of reducing the $(D + 2)$-dimensional super curvature
$\tilde F$ to the $D$-dimensional ordinary curvature $F$ in
the horizontality restriction ($\tilde F = F$) leads to (i) the derivation
of the nilpotent (anti-)BRST symmetry transformations on the gauge field and
the (anti-)ghost fields of the $p$-form gauge theory, (ii) 
the geometrical interpretation for the (anti-)BRST charges
$Q_{(a)b}$ as the translation generators along the Grassmannian directions
of the supermanifold, and (iii) the geometrical meaning of the
nilpotency ($s_{(a)b}^2 = 0, Q_{(a)b}^2 = 0$)
property which is found to be encoded in a
couple of successive translations (i.e. $(\partial/\partial\theta)^2 =
(\partial/\partial\bar\theta)^2 = 0$) along any particular Grassmannian
direction (i.e. $\theta$ or $\bar\theta$) of the supermanifold.

Recently, in a set of papers [24-26], all the three super cohomological
operators $(\tilde d, \tilde \delta, \tilde \Delta)$ corresponding to
the ordinary de Rham cohomological operators
\footnote{The set $(d, \delta, \Delta)$ of operators,
defined on a compact manifold without a boundary, is called the set
of de Rham cohomological operators where $
\delta = \pm * d *, d = dx^\mu \partial_\mu,
\Delta = (d + \delta)^2 $ are called the (co-)exterior
derivatives ($(\delta)d$) and the Laplacian operator 
($\Delta$) respectively. Here $*$ is the Hodge 
duality operation on the manifold. These operators obey an algebra:
$d^2 = \delta^2 = 0, \Delta = \{ d, \delta \}, [\delta, \Delta] = 0,
[d, \Delta] = 0$ showing that the Laplacian
operator $\Delta$ is the Casimir operator for the
whole algebra (see, eg, [17,18] for details).} 
have been exploited in the generalized versions of the horizontality 
condition defined on a four $( 2 + 2)$-dimensional supermanifold
to demonstrate the existence of the local, covariant and continuous
(anti-)BRST-, (anti-)co-BRST- and a bosonic symmetry
(which is equal to the anticommutator(s) of the (anti-)BRST and (anti-)co-BRST
symmetries) transformations for the two $(1 + 1)$-dimensional (2D) free 
Abelian gauge theory. The above symmetry transformations have also been 
discussed in the canonical Lagrangian formulation of this theory [27-29].
Exactly similar kind of symmetry transformations for the 
self-interacting 2D non-Abelian gauge theories have also been obtained in
the Lagrangian formulation [30] as  well as superfield formulation [31]. The
topological nature of the above 2D (non-)Abelian gauge theories has also 
been captured in the superfield formulation where, for the first time,
the geometrical origin for the Lagrangian density and the symmetric energy
momentum tensor has been provided. In fact,
these physical quantities have been shown
to correspond to the translations of some local (but composite) superfields
along the Grassmannian directions of the supermanifold [32,33]. In a very 
recent paper [34], the local, covariant, continuous and nilpotent (anti-)BRST
symmetry transformations and the non-local, non-covariant, continuous and
nilpotent (anti-)co-BRST transformations have been shown to exist in
the superfield formulation for the 4D interacting 
Abelian gauge theory defined on the six $(4 + 2)$-dimensional supermanifold. 
For the 4D free Abelian 2-form gauge theory, the local, covariant, continuous
and nilpotent (anti-)BRST and (anti-)co-BRST symmetries as well as a bosonic
symmetry have also been obtained in the Lagrangian formulation [35,36].

In all the above-cited papers on the superfield formalism, only the nilpotent
transformations for the gauge field
and the (anti-)ghost fields have been obtained
by exploiting the (dual-)horizontality conditions on the supermanifolds
(see, eg, [26,34] for detail references).
The horizontality condition ($\tilde F = F$) and the dual-horizontality
condition ($\tilde d \tilde A = \delta A$) owe their origin to (i) the
super (co-)exterior derivatives $(\tilde \delta)\tilde d$ and their
ordinary counterparts $(\delta)d$, and (ii) the super 1-form connection
$\tilde A$ and its ordinary counterpart $A$. In physical terms, the above
conditions originate due to the gauge (or BRST) invariance of the
$(p + 1)$-form (super)curvatures $(\tilde F)F$ and the dual-gauge
(or co-BRST) invariance of the (super) zero-forms $\tilde \delta \tilde A$
and $\delta A$, respectively. It is obvious that, in the above conditions
on the (super)manifolds,
the matter fields of the interacting gauge theory {\it do not}
 play any role at all. As a consequence, these conditions 
do not shed any light on the derivation of the nilpotent 
symmetry transformations for the matter fields of the theory.
To the best of our knowledge, in the known literature on the superfield 
formulations [9-14, 24-26], there has been no definite clue on 
the derivation of the nilpotent symmetry transformations for the
matter fields. This is why, it has been a long-standing problem
to derive the nilpotent transformations for the matter fields for an 
{\it interacting} gauge theory in any arbitrary dimension of spacetime.
In this connection, it is worthwhile to mention that, in a very recent paper
[37], it has been shown that the invariance of the conserved 
matter (super)currents on the four $(2 + 2)$-dimensional (super)manifold
leads to the derivation of the nilpotent (anti-)BRST transformations
for the Dirac fields in an interacting 2D Abelian gauge theory where the
matter conserved current $J_\mu^{(d)} = \bar \psi \gamma_\mu \psi$ couples
to the $U(1)$ gauge field $A_\mu$. For the massless Dirac fields, it has been
shown that the invariance of the (super) axial-vector current,
constructed by the (super) matter fields, leads to the derivation of the local,
covariant, continuous and off-shell nilpotent (anti-)co-BRST transformations 
on the massless Dirac fields.

The purpose of the present paper is to demonstrate
that the invariance of the vector conserved (super)currents
(i.e. $\tilde J_\mu^{(d,c)} (x,\theta,\bar\theta) = J_\mu^{(d,c)} (x)$), 
constructed by the (super) Dirac and (super) complex scalar fields, on the six 
$(4 + 2)$-dimensional supermanifold leads to the derivation of the off-shell
nilpotent, local, covariant and continuous (anti-)BRST symmetries for the 
Dirac- as well as complex scalar fields. We would like to lay emphasis on
the fact that, the requirement of the invariance of the matter (super)currents,
is not a restriction put by hand from outside. Rather, it is the inherent 
and innate feature of the interacting
gauge theory itself. Thus, this condition emerges automatically, unlike the
case of the (dual-)horizontality conditions (see, eg, [26] for details)
which are imposed  by hand on the supermanifold. For the case of the
interacting $U(1)$ gauge theory with the Dirac fields, we show that the
horizontality condition does not play any significant role in the derivation 
of the nilpotent (anti-)BRST symmetries for the Dirac fields. This is because 
of the fact that the matter conserved current 
$J_\mu^{(d)} = \bar \psi \gamma_\mu \psi$ does not contain, in any way,
the other physical field $A_\mu$ of the theory. On the contrary, in the case of 
the interacting Abelian gauge theory involving the complex scalar 
fields, the horizontality condition does play a very important role in the
derivation of the nilpotent symmetries on the matter (complex scalar) fields.
The root cause of this crucial role, played by the horizontality condition,
is the presence of the gauge field $A_\mu$ in the conserved matter current 
$J_\mu^{(c)} \sim \phi^* \partial_\mu \phi - \phi \partial_\mu \phi^*
+ 2 i e A_\mu \phi^* \phi$ constructed by the complex scalar fields.
In fact, the interacting $U(1)$ gauge theory with the complex scalar field
provides a really interesting physical system where the
horizontality condition and the invariance of the conserved matter 
(super)currents on the (super)manifolds are found to be consistent with
each-other. This mutual consistency entails upon the nilpotent transformations 
for the gauge field, the (anti-)ghost 
fields and the matter fields to be complementary
to one-another. We comment more on this {\it consistency issue}
in the conclusion (cf section 6) part of our present paper.

The contents of our present paper are organized as follows. In section 2, we
give a brief synopsis of the off-shell nilpotent
(anti-)BRST symmetries for the interacting
$U(1)$ gauge theory in the Lagrangian formulation
where the gauge field $A_\mu$ is coupled to the 
conserved matter currents constructed by (i) the Dirac fields, and
(ii) the complex scalar fields. For the sake of this paper to be 
self-contained, section 3 deals with the derivation of
the above nilpotent symmetries for the gauge- and (anti-)ghost fields
in the framework of superfield formulation where the horizontality
condition on the six $(4 + 2)$-dimensional supermanifold plays a
crucial role [12,26]. The central theme of our paper is
contained in  sections 4 and 5 where
we derive the off-shell nilpotent symmetries for the Dirac and complex scalar
fields, respectively, by exploiting the invariance of the conserved
matter (super)currents on the (super)manifolds. We lay emphasis on our key
results, make some concluding remarks and point out a few future directions for
further investigations in section 6.\\

\noindent
{\bf 2 Nilpotent (anti-)BRST symmetries: Lagrangian formulation}\\

\noindent
To recapitulate the key points connected 
with the local, covariant, continuous and off-shell nilpotent
(anti-)BRST symmetries for the Lagrangian density ${\cal L}_{b}$
of an {\it interacting} four ($3 + 1)$-dimensional (4D) $U(1)$ gauge theory
\footnote{We adopt here the conventions and notations such that the 4D flat
Minkowski metric is: $\eta_{\mu\nu} =$ diag $(+1, -1, -1, -1)$ and $\Box = 
\eta^{\mu\nu} \partial_{\mu} \partial_{\nu} = (\partial_{0})^2 - 
(\partial_{i})^2, F_{0i} 
= \partial_{0} A_{i} - \partial_{i} A_{0} = E_i \equiv {\bf E},
F_{ij} = \epsilon_{ijk} B_k, B_i \equiv  {\bf B} = \frac{1}{2} \epsilon_{ijk}
F_{jk}, (\partial \cdot  A) = \partial_{0} A_0 - \partial_i A_i, 
D_{\mu} \psi = \partial_{\mu} \psi + i e 
A_{\mu} \psi$ where ${\bf E}$ and ${\bf B}$ are the electric and magnetic 
fields, respectively. The totally antisymmetric Levi-Civita tensor 
$\varepsilon_{\mu\nu\lambda\xi}$ and
the $4 \times 4$ Dirac $\gamma$-matrices are chosen to satisfy:
$ \varepsilon_{0123} = -\varepsilon^{0123} = + 1, \; \varepsilon_{0ijk}
= \epsilon_{ijk}, \{ \gamma^\mu, \gamma^\nu \} = 2 \eta^{\mu\nu}$
where $\epsilon_{ijk}$ is the Levi-Civita tensor in the space submanifold. 
Here the Greek indices: $\mu, \nu, \lambda...
= 0, 1, 2, 3$ correspond to the spacetime directions 
and Latin indices $i, j, k, ...= 1, 2, 3$ stand only for the 
space directions on the manifold.} 
in the Feynman gauge, we begin with [38-40]
$$
\begin{array}{lcl}
{\cal L}_{b} &=& - \frac{1}{4}\; F^{\mu\nu} F_{\mu\nu} 
+ \bar \psi \;(i \gamma^\mu D_\mu - m)\; \psi + B \;(\partial \cdot A)
+ \frac{1}{2}\; B^2
- i \;\partial_{\mu} \bar C \partial^\mu C \nonumber\\
&\equiv& \frac{1}{2}\; ({\bf E^2} - {\bf B^2})
+ \bar \psi\; (i \gamma^\mu D_\mu - m) \;\psi + B \;(\partial \cdot A)
+ \frac{1}{2}\; B^2
- i \;\partial_{\mu} \bar C \partial^\mu C 
\end{array} \eqno(2.1)
$$
where $F_{\mu\nu} = \partial_\mu A_\nu - \partial_\nu A_\mu$ is the field 
strength tensor for the $U(1)$ gauge field $A_\mu$ that is
derived from the 2-form $d A = \frac{1}{2} (dx^\mu \wedge dx^\nu) F_{\mu\nu}$. 
As is evident, the latter
is  constructed by the application of the exterior derivative 
$d = dx^\mu \partial_\mu$ (with $d^2 = 0)$ on the 1-form $A = dx^\mu A_\mu$
(which defines the vector potential $A_\mu$). The
gauge-fixing term $(\partial \cdot A)$ is derived through the operation
of the co-exterior derivative $\delta$ 
(with $\delta = - * d *, \delta^2 = 0$) on the
one-form $A$ (i.e. $\delta A = - * d * A = (\partial \cdot A)$)
where $*$ is the Hodge duality operation. The fermionic
Dirac fields $(\psi, \bar \psi)$, with mass $m$ and charge $e$, couple 
to the $U(1)$ gauge field $A_\mu$ (i.e. $ - e \bar \psi \gamma^\mu 
A_\mu \psi$) through the conserved current 
$J_\mu^{(d)} = \bar \psi \gamma_\mu \psi$. The
anticommuting ($ C \bar C + \bar C C = 0, C^2 = \bar C^2 = 0,
C \psi + \psi C = 0$ etc.) (anti-)ghost fields $(\bar C)C$ are required to
maintain both 
the unitarity and ``quantum'' gauge (i.e. BRST) invariance together
at any arbitrary order of perturbation theory
\footnote{ The full strength of the (anti-)ghost fields is realized in the
discussion of the unitarity and gauge invariance
for the perturbative computations in the realm of non-Abelian gauge theory
where the Feynman graphs involve
the loop diagrams of the gauge (gluon) fields
(see, eg, [22] for details). In fact, to maintain the unitarity, there
exists a ghost loop diagram corresponding to a loop diagram involving only the
gauge field. This is required to counter the contributions coming out
from the gauge loop graph [22].}. The Nakanishi-Lautrup auxiliary field 
$B = - (\partial \cdot A)$
is required to linearize the usual gauge-fixing term $- \frac{1}{2} 
(\partial \cdot A)^2$ of the theory. Under the following off-shell nilpotent
$(s_{(a)b}^2 = 0)$ (anti-)BRST symmetry transformations $s_{(a)b}$
\footnote{We adopt here
the notations and conventions followed by Weinberg [40]. In its 
totality, the nilpotent ($\delta_B^2 = 0$)
BRST transformation $\delta_B$ is the product of an
anticommuting (i.e. $\eta C + C \eta = 0, \eta \psi + \psi \eta = 0$ etc.)
spacetime independent parameter $\eta$ and $s_b$ as
$\delta_B = \eta s_{b}$ where $s_b^2 = 0$.}
on the gauge-, (anti-)ghost- and matter fields 
(with $s_{b} s_{ab} + s_{ab} s_b = 0$) [38-40]
$$
\begin{array}{lcl}
s_{b} A_{\mu} &=& \partial_{\mu} C\; \qquad 
s_{b} C = 0\; \qquad 
s_{b} \bar C = i B\;  \qquad s_b \psi = - i e C \psi \nonumber\\
s_b \bar \psi &=& - i e \bar \psi C\;
\qquad s_{b} {\bf B} = 0\; \quad  s_{b} B = 0\; \quad
\;s_{b} {\bf E} = 0\; \quad s_b (\partial \cdot A) = \Box C \nonumber\\
s_{ab} A_{\mu} &=& \partial_{\mu} \bar C\; \qquad 
s_{ab} \bar C = 0\; \qquad 
s_{ab} C = - i B\;  \qquad s_{ab} \psi = - i e \bar C \psi \nonumber\\
s_{ab} \bar \psi &=& - i e \bar \psi \bar C\;
\qquad s_{ab} {\bf B} = 0\; \quad  s_{ab} B = 0\; \quad
\;s_{ab} {\bf E} = 0\; \quad s_{ab} (\partial \cdot A) = \Box \bar C
\end{array}\eqno(2.2)
$$
the above Lagrangian density transforms to a total derivative. The above
transformations are generated by the  off-shell nilpotent ($Q_{(a)b}^2 = 0$)
and conserved (anti-)BRST charges $Q_{(a)b}$.

The other dynamically closed system 
\footnote{ In the sense of the basic requirements of
a canonical field theory, the Lagrangian density ${\cal L}_B$ (cf (2.3))
describes a dynamically closed system because the quadratic kinetic energy 
terms and the interaction terms for all the fields $\phi, \phi^*$ and $A_\mu$
are present in this Lagrangian
density in a logical fashion [41].} that
respects the above kind of symmetry transformations is the system of complex
scalar fields coupled to the $U(1)$ gauge field $A_\mu$. This system
is described by the following Lagrangian density (see, eg, [41])
$$
\begin{array}{lcl}
{\cal L}_{B} &=& - \frac{1}{4}\; F^{\mu\nu} F_{\mu\nu} 
+ (D_\mu \phi)^{*} D^\mu \phi - V (\phi^* \phi) + B \;(\partial \cdot A)
+ \frac{1}{2}\; B^2
- i \;\partial_{\mu} \bar C \partial^\mu C \nonumber\\
&\equiv& \frac{1}{2}\; ({\bf E^2} - {\bf B^2})
+ (D_\mu \phi)^{*} D^\mu \phi - V (\phi^* \phi) + B \;(\partial \cdot A)
+ \frac{1}{2}\; B^2
- i \;\partial_{\mu} \bar C \partial^\mu C
\end{array} \eqno(2.3)
$$
where $V(\phi^*\phi)$
\footnote{ For a renormalizable quantum field theory, this potential can
be chosen in the quartic polynomial form as:
$V(\phi^*\phi) = \mu^2 \phi^* \phi + \lambda (\phi^*\phi)^2$ where $\mu$
and $\lambda$ are the parameters which could be chosen in different ways
for different purposes. For instance, the free field theory corresponds to
$\lambda = 0, \mu^2 > 0$ (see, eg, [41] for details). The key point to
be noted here is the fact that $V(\phi^*\phi)$ remains invariant under (2.6).}
is the potential describing the interaction between
the complex scalar fields $\phi$ and $\phi^*$ and the covariant derivatives
on these fields are
$$
\begin{array}{lcl}
D_\mu \phi = \partial_\mu \phi + i e A_\mu \phi\; \qquad\;\;
(D_\mu \phi)^* = \partial_\mu \phi^* - i e A_\mu \phi^*.
\end{array} \eqno(2.4)
$$
It will be noted that the gauge field $A_\mu$ couples to 
the conserved matter current $J_\mu^{(c)} \sim [\phi^* D_\mu \phi -
\phi (D_\mu \phi)^*]$ to provide the interaction between the $U(1)$
gauge field and matter fields $\phi$ and $\phi^*$ (cf (2.3)).
This statement can be
succinctly expressed by re-expressing (2.3), in terms of the kinetic
energy terms for $\phi$ and $\phi^*$, as
$$
\begin{array}{lcl}
{\cal L}_{B} &=& - \frac{1}{4}\; F^{\mu\nu} F_{\mu\nu} 
+ \partial_\mu \phi^{*} \partial^\mu \phi - i e A_\mu [\phi^* \partial_\mu \phi
- \phi \partial_\mu \phi^*] + e^2 A^2 \phi^* \phi\nonumber\\
&-& V (\phi^* \phi) + B \;(\partial \cdot A)
+ \frac{1}{2}\; B^2
- i \;\partial_{\mu} \bar C \partial^\mu C.
\end{array} \eqno(2.5)
$$
The conservation ($\partial \cdot J^{(c)} = 0$)
of the matter current $J_\mu^{(c)}$ (which couples to the gauge field $A_\mu$
in the above Lagrangian density)
can be easily checked by exploiting the equations of motion
$D_\mu D^\mu \phi = - (\partial V/\partial \phi^*), 
(D_\mu D^\mu \phi)^*  = - (\partial V/\partial \phi)$ derived from
the Lagrangian densities (2.3) and/or (2.5). 
The above Lagrangian density respects the following off-shell 
nilpotent (anti-)BRST
transformations on the matter fields, gauge field and the (anti-)ghost fields:
$$
\begin{array}{lcl}
s_{b} A_{\mu} &=& \partial_{\mu} C\; \qquad 
s_{b} C = 0\; \qquad 
s_{b} \bar C = i B\;  \qquad s_b \phi = - i e C \phi \nonumber\\
s_b  \phi^* &=& + i e  \phi^* C\;
\qquad s_{b} {\bf B} = 0\; \quad  s_{b} B = 0\; \quad
\;s_{b} {\bf E} = 0\; \quad s_b (\partial \cdot A) = \Box C \nonumber\\
s_{ab} A_{\mu} &=& \partial_{\mu} \bar C\; \qquad 
s_{ab} \bar C = 0\; \qquad 
s_{ab} C = - i B\;  \qquad s_{ab} \phi = - i e \bar C \phi \nonumber\\
s_{ab} \phi^* &=& + i e  \phi^* \bar C\;
\qquad s_{ab} {\bf B} = 0\; \quad  s_{ab} B = 0\; \quad
\;s_{ab} {\bf E} = 0\; \quad s_{ab} (\partial \cdot A) = \Box \bar C. 
\end{array}\eqno(2.6)
$$
The key points to be noted, at this stage, are (i) under
the (anti-)BRST transformations, it is the kinetic energy term
$- \frac{1}{4} F^{\mu\nu} F_{\mu\nu}$ that remains invariant. This statement 
is true for any gauge theory. For the above $U(1)$ gauge theory, as it turns
out, it is the curvature term $F_{\mu\nu}$ (constructed from the operation
of the exterior derivative $d$ on the 1-form $A = dx^\mu A_\mu$)
itself that remains invariant under the (anti-)BRST transformations. (ii) In 
the mathematical language, the (anti-)BRST symmetries owe their origin to
the exterior derivative $d$ because the curvature term
is constructed from it. (iii) The gauge- and (anti-)ghost fields are endowed 
with exactly the same symmetry transformations for both the cases of the
interacting Abelian gauge theories that are being considered here (cf (2.2)
and (2.6)). (iv) In general, the above transformations can be concisely 
expressed in terms of the generic field $\Sigma (x)$ and conserved charges 
$Q_{(a)b}$, as
$$
\begin{array}{lcl}
s_{r}\; \Sigma (x) = - i\; 
\bigl [\; \Sigma (x),  Q_r\; \bigr ]_{\pm} \;\;\qquad\;\;\;
r = b, ab 
\end{array} \eqno(2.7)
$$
where the local generic field
$\Sigma = A_\mu, C, \bar C, \psi, \bar \psi, B,\phi, \phi^*$ 
and the $(+)-$ signs, 
as the subscripts on the (anti-)commutators $[\;, \;]_{\pm}$, 
stand for $\Sigma$ being (fermionic)bosonic in nature.\\

\noindent
{\bf 3 The gauge- and (anti-)ghost fields: nilpotent symmetries}\\

\noindent
We exploit here the superfield formalism to obtain the  
off-shell nilpotent symmetry transformations for $A_\mu, C, \bar C$ fields
present in (2.2) and/or (2.6). To this end in mind, we
begin with a six $(4 + 2)$-dimensional supermanifold parametrized by
the general superspace coordinate $Z^M = (x^\mu, \theta, \bar \theta)$
where $x^\mu (\mu = 0, 1, 2, 3)$ are the four even spacetime coordinates
and $\theta, \bar \theta$  are a couple of odd elements of a Grassmann
algebra. On this supermanifold, one can define a supervector superfield
$\tilde A_M = (B_\mu (x,\theta,\bar\theta), {\cal F} (x,\theta,\bar\theta),
\bar {\cal F} (x,\theta,\bar\theta))$ 
with $B_\mu, {\cal F}, \bar {\cal F}$ as the component
multiplet superfields [12,11]. These multiplet superfields can be 
expanded in terms of the basic fields $A_\mu, C, \bar C$ and the secondary
fields as (see, eg, [11,12,26])
$$
\begin{array}{lcl}
B_{\mu} (x, \theta, \bar \theta) &=& A_{\mu} (x) 
+ \theta\; \bar R_{\mu} (x) + \bar \theta\; R_{\mu} (x) 
+ i \;\theta \;\bar \theta S_{\mu} (x) \nonumber\\
{\cal F} (x, \theta, \bar \theta) &=& C (x) 
+ i\; \theta \bar B (x)
+ i \;\bar \theta\; {\cal B} (x) 
+ i\; \theta\; \bar \theta \;s (x) \nonumber\\
\bar {\cal F}  (x, \theta, \bar \theta) &=& \bar C (x) 
+ i \;\theta\;\bar {\cal B} (x) + i\; \bar \theta \;B (x) 
+ i \;\theta \;\bar \theta \;\bar s (x).
\end{array} \eqno(3.1)
$$
It is straightforward to note that the local 
fields $ R_{\mu} (x), \bar R_{\mu} (x),
C (x), \bar C (x), s (x), \bar s (x)$ are fermionic (anti-commuting) 
in nature and the bosonic (commuting) local fields in (3.1)
are: $A_{\mu} (x), S_{\mu} (x), {\cal B} (x), \bar {\cal B} (x),
B (x), \bar B (x)$. It is clear
that, in the above expansion, the bosonic-
 and fermionic degrees of freedom match and, in the limit
$\theta, \bar\theta \rightarrow 0$, we get back our basic
fields $A_\mu, C, \bar C$ of (2.1) and/or (2.5). These requirements 
are essential
for the sanctity of any arbitrary supersymmetric theory in the 
superfield formulation. In fact, all the secondary fields will be expressed 
in terms of basic fields due to the restrictions emerging from the application 
of horizontality condition (i.e. $\tilde F = F$), namely;
$$
\begin{array}{lcl} 
\tilde F =  \frac{1}{2}\; (d Z^M \wedge d Z^N)\;
\tilde F_{MN} = \tilde d \tilde A  \equiv
d A = \frac{1}{2} (dx^\mu \wedge dx^\nu)\; F_{\mu\nu} = F
\end{array} \eqno(3.2)
$$
where the super exterior derivative $\tilde d$ and 
the connection super one-form $\tilde A$ are defined as
$$
\begin{array}{lcl}
\tilde d &=& \;d Z^M \;\partial_{M} = d x^\mu\; \partial_\mu\;
+ \;d \theta \;\partial_{\theta}\; + \;d \bar \theta \;\partial_{\bar \theta}
\nonumber\\
\tilde A &=& d Z^M\; \tilde A_{M} = d x^\mu \;B_{\mu} (x , \theta, \bar \theta)
+ d \theta\; \bar {\cal F} (x, \theta, \bar \theta) + d \bar \theta\;
{\cal F} ( x, \theta, \bar \theta).
\end{array}\eqno(3.3)
$$
In physical language, the requirement (3.2) implies that the physical field
${\bf E}$ and ${\bf B}$, derived from the curvature term $F_{\mu\nu}$, 
do not get any
contribution from the Grassmannian variables. In other words, the
physical electric field ${\bf E}$ and magnetic field
${\bf B}$ for the 4D QED remain intact in the
superfield formulation. Mathematically, the condition (3.2) implies
the ``flatness'' of all the components of the
super curvature (2-form) tensor $\tilde F_{MN}$ that are directed along the 
 $\theta$ and/or $\bar \theta$ directions of the supermanifold. To this
end in mind, first we expand $\tilde d \tilde A = \tilde F$ (for the
interacting Abelian gauge theory under consideration) as
$$
\begin{array}{lcl}
\tilde d \tilde A &=& (d x^\mu \wedge d x^\nu)\;
(\partial_{\mu} B_\nu) - (d \theta \wedge d \theta)\; (\partial_{\theta}
\bar {\cal F}) + (d x^\mu \wedge d \bar \theta)
(\partial_{\mu} {\cal F} - \partial_{\bar \theta} B_{\mu}) \nonumber\\
&-& (d \theta \wedge d \bar \theta) (\partial_{\theta} {\cal F}
+ \partial_{\bar \theta} \bar {\cal F}) 
+ (d x^\mu \wedge d \theta) (\partial_{\mu} \bar {\cal F} - \partial_{\theta}
B_{\mu}) - (d \bar \theta \wedge d \bar \theta)
(\partial_{\bar \theta} {\cal F}). 
\end{array}\eqno(3.4)
$$
Ultimately, the application of soul-flatness (horizontality) condition
($\tilde d \tilde A = d A$) yields [26]
$$
\begin{array}{lcl}
R_{\mu} \;(x) &=& \partial_{\mu}\; C(x)\; \qquad 
\bar R_{\mu}\; (x) = \partial_{\mu}\;
\bar C (x)\; \qquad \;s\; (x) = \bar s\; (x) = 0
\nonumber\\
S_{\mu}\; (x) &=& \partial_{\mu} B\; (x) 
\qquad\;
B\; (x) + \bar B \;(x) = 0\; \qquad 
{\cal B}\; (x)  = \bar {\cal B} (x) = 0.
\end{array} \eqno(3.5)
$$
The insertion of all the above values in the expansion (3.1) yields
$$
\begin{array}{lcl}
B_{\mu} (x, \theta, \bar \theta) &=& A_{\mu} (x) 
+ \theta\; \partial_{\mu} \bar C (x) + \bar \theta\; \partial_{\mu} C (x) 
+ i \;\theta \;\bar \theta \partial_{\mu} B (x) \nonumber\\
{\cal F} (x, \theta, \bar \theta) &=& C (x) 
- i\; \theta B (x)\;
\qquad\;
\bar {\cal F}  (x, \theta, \bar \theta) = \bar C (x) 
+ i\; \bar \theta \;B (x).
\end{array} \eqno(3.6)
$$
This equation leads to
the derivation of the (anti-)BRST symmetries for the 
gauge- and (anti-)ghost fields of the Abelian gauge theory (cf (2.2)
and  (2.6)).
In addition, this exercise provides  the physical interpretation for the
(anti-)BRST charges $Q_{(a)b}$ 
as the generators (cf eqn. (2.7)) of translations 
(i.e. $ \mbox{Lim}_{\bar\theta \rightarrow 0} (\partial/\partial \theta),
 \mbox{Lim}_{\theta \rightarrow 0} (\partial/\partial \bar\theta)$)
along the Grassmannian
directions of the supermanifold. Both these observations can be succinctly 
expressed, in a combined fashion, by re-writing the super expansion (3.1) as
$$
\begin{array}{lcl}
B_{\mu}\; (x, \theta, \bar \theta) &=& A_{\mu} (x) 
+ \;\theta\; (s_{ab} A_{\mu} (x)) 
+ \;\bar \theta\; (s_{b} A_{\mu} (x)) 
+ \;\theta \;\bar \theta \;(s_{b} s_{ab} A_{\mu} (x)) \nonumber\\
{\cal F}\; (x, \theta, \bar \theta) &=& C (x) \;+ \; \theta\; (s_{ab} C (x))
\;+ \;\bar \theta\; (s_{b} C (x)) 
\;+ \;\theta \;\bar \theta \;(s_{b}\; s_{ab} C (x))
 \nonumber\\
\bar {\cal F}\; (x, \theta, \bar \theta) &=& \bar C (x) 
\;+ \;\theta\;(s_{ab} \bar C (x)) \;+\bar \theta\; (s_{b} \bar C (x))
\;+\;\theta\;\bar \theta \;(s_{b} \;s_{ab} \bar C (x)).
\end{array} \eqno(3.7)
$$
A closer look at the (anti-)BRST transformations shows that the (anti-)ghost
fields transform only under one of these transformations. That is to say
the fact that the anti-ghost field $\bar C$ transforms under the BRST
transformation but it remains unchanged under the anti-BRST transformation.
Exactly the opposite happens with the ghost field $C$. This statement can
be expressed in a more sophisticated language of the 
conditions on the superfields. It is
clear from (3.6) that the horizontality condition enforces the superfields
$(\bar {\cal F} (x,\theta,\bar\theta)) {\cal F} (x,\theta,\bar\theta)$ to 
become (anti-)chiral due to the equivalence between the translation generators
operating on superfields of the supermanifold and the internal symmetry 
generators $Q_{(a)b}$ acting on the local fields of the ordinary manifold
(cf (4.10) below).\\

\noindent
{\bf 4 The Dirac fields: nilpotent symmetries}\\

\noindent
It is clear that, for the derivation of the off-shell nilpotent
symmetries on the gauge- and (anti-)ghost fields, we do require the
horizontality restriction on the supermanifold. For this purpose, the basic 
physical
and mathematical objects we exploit, are the connection (super) one-form
$(\tilde A) A$ and the (super) exterior derivative $(\tilde d) d$ to
obtain the symmetry transformations on the above fields (by the
restriction $\tilde F = F$). Consistent with the above nilpotent 
transformations, the nilpotent transformations on the Dirac fields
are derived from the requirement of the invariance of the matter
conserved (super)currents on the (super)manifolds. To corroborate this
assertion, we begin with the super expansions for the Dirac superfields
$\Psi (x,\theta, \bar\theta)$ and $\bar \Psi (x,\theta,\bar\theta)$
in terms of the basic Dirac fields $\psi (x)$ and $\bar\psi (x)$
and some extra secondary fields as
$$
\begin{array}{lcl} 
 \Psi (x, \theta, \bar\theta) &=& \psi (x)
+ i \;\theta\; \bar b_1 (x) + i \;\bar \theta \; b_2 (x) 
+ i \;\theta \;\bar \theta \;f (x)
\nonumber\\
\bar \Psi (x, \theta, \bar\theta) &=& \bar \psi (x)
+ i\; \theta \;\bar b_2 (x) + i \;\bar \theta \; b_1 (x) 
+ i\; \theta \;\bar \theta \;\bar f (x).
\end{array} \eqno(4.1)
$$
It is evident that, in the limit 
$(\theta, \bar\theta) \rightarrow 0$,
we get back the Dirac fields $(\psi, \bar\psi)$ of the 
Lagrangian density (2.1). Furthermore, the number of
bosonic fields ($b_1, \bar b_1, b_2, \bar b_2)$ match with the fermionic
fields $(\psi, \bar \psi, f, \bar f)$ so that the above expansion is consistent
with the basic tenets of supersymmetry. Now one can construct the
supercurrent $\tilde J_\mu (x, \theta, \bar\theta)$ from the above
superfields with the following general super expansion
$$
\begin{array}{lcl} 
\tilde J_\mu^{(d)}\; (x, \theta, \bar\theta) &=& \bar \Psi (x,\theta,\bar\theta)
\;\gamma_\mu \;\Psi (x, \theta, \bar\theta) \nonumber\\
&=& J_\mu^{(d)}\; (x) + \theta \; \bar K_\mu^{(d)}\; (x)
+ \bar \theta \;K_\mu^{(d)}\; (x) 
+ i\; \theta\; \bar\theta\; L_\mu^{(d)}\; (x) 
\end{array} \eqno(4.2)
$$
where the above components (i.e. $\bar K_\mu^{(d)}, 
K_\mu^{(d)}, L_\mu^{(d)}, J_\mu^{(d)}$),
along the  Grassmannian directions
$\theta$ and  $\bar\theta$ as well as the bosonic directions
$\theta\bar\theta$ and identity $\hat {\bf 1}$ of the
supermanifold, can be expressed in terms of the components of the
basic super expansions (4.1), as
$$
\begin{array}{lcl} 
&& \bar K_\mu^{(d)} (x) = i \bigl ( \bar b_2 \gamma_\mu \psi -
\bar \psi \gamma_\mu \bar b_1 \bigr )\;
\qquad \; K_\mu^{(d)} (x) = i \bigl ( b_1 \gamma_\mu \psi -
\bar \psi \gamma_\mu  b_2 \bigr ) \nonumber\\
&& L_\mu^{(d)} (x) = \bar f \gamma_\mu \psi + \bar \psi \gamma_\mu f
+ i (\bar b_2 \gamma_\mu b_2 - b_1 \gamma_\mu \bar b_1)\; 
\qquad\; J_\mu^{(d)} (x) = \bar \psi  \gamma_\mu \psi.
 \end{array} \eqno(4.3)
$$
To be consistent with our earlier observation that the (anti-)BRST
transformations $(s_{(a)b})$ are equivalent to the translations
($\mbox{Lim}_{\bar\theta \rightarrow 0} (\partial/\partial \theta)$)
$\mbox{Lim}_{\theta \rightarrow 0} (\partial/\partial \bar\theta)$
along the $(\theta)\bar\theta$-directions of the supermanifold, it is 
straightforward
to re-express the expansion in (4.2) as follows
$$
\begin{array}{lcl} 
\tilde J_\mu^{(d)} (x, \theta, \bar\theta) = J_\mu^{(d)} (x) + \theta \; 
(s_{ab} J_\mu^{(d)} (x)) + \bar \theta\; (s_b J_\mu^{(d)} (x)) 
+ \theta\; \bar\theta\; (s_b s_{ab} J_\mu^{(d)} (x)).
\end{array} \eqno(4.4)
$$
It can be checked that, under the (anti-)BRST transformations (2.2),
the conserved current $J_\mu^{(d)} (x)$ remains invariant
(i.e. $s_{b} J_\mu^{(d)} (x) = s_{ab} J_\mu^{(d)} (x) = 0$).
This statement, with the help of (4.2) and (4.4),
 can be mathematically expressed as
$$
\begin{array}{lcl} 
&&s_b J_\mu^{(d)} = 0 \Rightarrow K_\mu^{(d)}
 = 0 \Rightarrow b_1 \gamma_\mu \psi
= \bar \psi \gamma_\mu b_2 \nonumber\\
&& s_{ab} J_\mu^{(d)} = 0 \Rightarrow \bar K_\mu^{(d)} = 0 \Rightarrow 
\bar b_2 \gamma_\mu \psi
= \bar \psi \gamma_\mu \bar b_1 \nonumber\\
&&s_b \bar s_{ab} J_\mu^{(d)} = 0 \Rightarrow L_\mu^{(d)} = 0 \Rightarrow 
\bar f \gamma_\mu \psi
+ \bar \psi \gamma_\mu f = i (b_1 \gamma_\mu \bar b_1 - \bar b_2
\gamma_\mu b_2). 
\end{array} \eqno(4.5)
$$
One of the possible solutions of the above restrictions, in terms
of the components  of the basic expansions in (4.1) and the
basic fields of the Lagrangian density (2.1), is
$$
\begin{array}{lcl}
&& b_1 = - e \bar \psi C\; \qquad b_2 = - e C \psi\;
\qquad \bar b_1 = - e \bar C \psi\; \qquad \bar b_2 = - e \bar \psi \bar C
\nonumber\\
&& f = - i e\; [\; B + e \bar C C\; ]\; \psi\;\;
\qquad \;\;\bar f = + i e\; \bar \psi\; [\; B + e C \bar C \;].
\end{array} \eqno(4.6)
$$
At the moment, it appears to us that the above solutions are the
{\it unique} solutions to all the restrictions in (4.5)
\footnote{ Let us concentrate on $b_1 \gamma_\mu \psi = \bar\psi
\gamma_\mu b_2$. A closer look at it makes it evident that the pair
of bosonic components $b_1$ and $b_2$ should be proportional
to the pair of fermionic fields $\bar\psi$ and $\psi$, respectively.
To make the latter pair bosonic in nature, we have to include the ghost
field $C$ of the Lagrangian
density (2.1) to obtain: $b_1 \sim \bar\psi C, b_2 \sim C \psi$.
Rest of the choices in (4.6) follow exactly similar kind of arguments.}. 
However, we do
not have a mathematically rigorous proof for the same.
Ultimately, the restriction that emerges on the $(2 + 2)$-dimensional
supermanifold is
$$
\begin{array}{lcl}
\tilde J_\mu^{(d)} (x, \theta, \bar \theta ) = J_\mu^{(d)} (x).
\end{array}\eqno (4.7)
$$
Physically, the above mathematical equation implies that there is
no superspace contribution to the ordinary conserved current 
$J_\mu^{(d)} (x)$. In
other words, the transformations on the Dirac fields $\psi$ and
$\bar\psi$ (cf (2.2)) are such that the supercurrent 
$\tilde J_\mu^{(d)} (x,\theta,\bar\theta)$ becomes a local composite field
$J_\mu^{(d)} (x) = (\bar\psi \gamma_\mu \psi) (x)$ 
{\it vis-{\'a}-vis} equation (4.4) and there is no Grassmannian
contribution to it. In a more sophisticated language, the
conservation law $\partial \cdot J^{(d)} = 0$ remains intact despite our 
discussions connected with the superspace and supersymmetry. It is
straightforward to check that the substitution of (4.6) into (4.1) 
leads to the following
$$
\begin{array}{lcl}
\Psi\; (x, \theta, \bar \theta) &=& \psi (x) \;+ \; \theta\; 
(s_{ab}  \psi (x))
\;+ \;\bar \theta\; (s_{b} \psi (x)) 
\;+ \;\theta \;\bar \theta \;(s_{b}\;  s_{ab} \psi (x))
 \nonumber\\
\bar \Psi\; (x, \theta, \bar \theta) &=& \bar \psi (x) 
\;+ \;\theta\;(s_{ab} \bar \psi (x)) \;+\bar \theta\; (s_{b} \bar \psi (x))
\;+\;\theta\;\bar \theta \;(s_{b} \; s_{ab} \bar \psi (x)).
\end{array} \eqno(4.8)
$$
This establishes the fact that the nilpotent (anti-)BRST charges $Q_{(a)b}$
are the translations generators 
$(\mbox{ Lim}_{\bar \theta \rightarrow 0}(\partial/\partial \theta))
\mbox{ Lim}_{ \theta \rightarrow 0}(\partial/\partial \bar \theta)$ 
along the $(\theta)\bar\theta$-directions
of the supermanifold. The property of the nilpotency 
(i.e. $Q_{(a)b}^2 = 0$) is encoded in the
two successive translations along the Grassmannian directions of the
supermanifold (i.e. $(\partial/\partial\theta)^2 = 
(\partial/\partial\bar\theta)^2 = 0$). In a more sophisticated
mathematical language, the above statement for the (anti-)BRST charges
$Q_{(a)b}$ can be succinctly expressed, using (2.7), as
$$
\begin{array}{lcl}
s_{b} \Sigma (x) &=& \mbox{Lim}_{\theta \rightarrow 0}\;
{\displaystyle \frac{\partial}{\partial \bar\theta}}\;
\tilde \Sigma (x,\theta,\bar\theta) \equiv - i \{\Sigma (x), Q_b \}
\nonumber\\
s_{ab} \Sigma (x) &=& \mbox{Lim}_{\bar\theta \rightarrow 0}\;
{\displaystyle \frac{\partial}{\partial \theta}}\;
\tilde \Sigma (x,\theta,\bar\theta) \equiv - i \{\Sigma (x), Q_{ab} \}
\end{array} \eqno(4.9)
$$
where the generic local field
$\Sigma (x) = \psi (x), \bar \psi (x)$ and the generic superfield
$\tilde \Sigma (x,\theta,\bar\theta) = \Psi (x,\theta,\bar\theta),
\bar \Psi (x,\theta,\bar\theta)$. Thus, it is evident that the 
nilpotent symmetry transformations, the corresponding nilpotent
charges and the translations generators on the supermanifold
are inter-related through the following mappings
$$
\begin{array}{lcl}
s_b \leftrightarrow Q_b \leftrightarrow
\mbox{Lim}_{\theta \rightarrow 0}\;
{\displaystyle \frac{\partial}{\partial \bar\theta}} \;\qquad\;
s_{ab} \leftrightarrow Q_{ab} \leftrightarrow
\mbox{Lim}_{\bar\theta \rightarrow 0}\;
{\displaystyle \frac{\partial}{\partial \theta}}.
\end{array} \eqno(4.10)
$$
The above relationship demonstrates that (i) the internal symmetry
transformations  on the ordinary fields, (ii) the nilpotent generators 
for the internal symmetry transformations, and (iii) 
the translation generators for the
superfields on the supermanifold are inextricably
intertwined with one-another.\\

\noindent
{\bf 5 The complex scalar fields: nilpotent symmetries}\\

\noindent
The central claim of our present investigation is connected with
our observation that the {\it invariance} of the (super)currents,
constructed by the (super) matter fields,
on the (super)manifolds leads to the derivation of the local, covariant,
continuous and off-shell nilpotent
(anti-)BRST symmetry transformations for the matter fields. 
In the previous section, we checked
the validity of the above claim in the context of the interacting
$U(1)$ gauge theory where the Dirac fields were coupled
to the $U(1)$ gauge field $A_\mu$. In this context, it is crucial to note
that both the conditions (i.e. horizontality restriction- and the 
invariance of the conserved currents on the supermanifold) are not
connected  with 
each-other in the case of interacting gauge theory with Dirac fields. 
These conditions are disjoint and decoupled in some sense. 
This is why, in the present section, we study the
complex scalar field coupled to the $U(1)$ gauge field which provides
an {\it interacting}
 system where the interplay between both the above restrictions plays
a crucial and decisive role
in the derivation of the off-shell nilpotent symmetries for the
matter fields. To bolster up this statement, we start off with the super 
expansion of the superfields $\Phi (x,\theta,\bar\theta)$ and 
$\Phi^* (x,\theta,\bar\theta)$ in terms of the basic fields $\phi (x)$
and $\phi^* (x)$ and some extra secondary fields, as
$$
\begin{array}{lcl} 
 \Phi (x, \theta, \bar\theta) &=& \phi (x)
+ i \;\theta\; \bar f_1 (x) + i \;\bar \theta \; f_2 (x) 
+ i \;\theta \;\bar \theta \;b (x)
\nonumber\\
 \Phi^* (x, \theta, \bar\theta) &=&  \phi^* (x)
+ i\; \theta \;\bar f^*_2 (x) + i \;\bar \theta \; f^*_1 (x) 
+ i\; \theta \;\bar \theta \; b^* (x)
\end{array} \eqno(5.1)
$$
where the number of fermionic local fields 
$\bar f_1 (x), f^*_1 (x), f_2 (x), \bar f^*_2 (x)$
match with the number of bosonic local
fields $\phi (x), \phi^* (x), b (x), b^* (x)$ to maintain
the basic requirements of a supersymmetric field theory. It is obvious
that, in the limit $(\theta, \bar \theta) \rightarrow 0$, we retrieve
our starting basic complex scalar
fields $\phi$ and $\phi^*$. In terms of the above superfields,
we can  write the expression for the supercurrent on the
supermanifold as
$$
\begin{array}{lcl} 
\tilde J_\mu^{(c)}\; (x, \theta, \bar\theta) &=& 
\Phi^* (x,\theta,\bar\theta) \;\partial_\mu\;
\Phi (x, \theta, \bar\theta)  
- \Phi (x,\theta,\bar\theta)\; \partial_\mu\;
\Phi^* (x, \theta, \bar\theta)  \nonumber\\
&+& 2\; i\; e\; B_\mu (x,\theta,\bar\theta)\; \Phi (x,\theta,\bar\theta) \;
\Phi^* (x,\theta,\bar\theta)
\end{array} \eqno(5.2)
$$
where $B_\mu (x,\theta,\bar\theta)$ is the superfield corresponding to
the vector $U(1)$ gauge field $A_\mu (x)$ that has the expansion (3.6).
It will be recalled that this expansion is obtained after the application of
the horizontality condition. The above supercurrent can be expanded, in
general, along the $\hat {\bf 1}, \theta, \bar\theta$
and $\theta\bar\theta$-directions of the supermanifold as
$$
\begin{array}{lcl} 
\tilde J_\mu^{(c)}\; (x, \theta, \bar\theta) 
= J_\mu^{(c)}\; (x) + \theta \; \bar K_\mu^{(c)}\; (x)
+ \bar \theta \;K_\mu^{(c)}\; (x) 
+ i\; \theta\; \bar\theta\; L_\mu^{(c)}\; (x)
\end{array} \eqno(5.3)
$$
where the individual components on the r.h.s can be expressed as follows
$$
\begin{array}{lcl}
 J^{(c)}_\mu (x) &=& \phi^* \partial_\mu \phi - \phi \partial_\mu \phi^*
+ 2 i e A_\mu \phi^* \phi \; \qquad
L_\mu^{(c)} (x) = L_{\mu 1}^{(c)} + L_{\mu 2}^{(c)}\nonumber\\
K_\mu^{(c)} &=& i \;\bigl [\; \phi^* \partial_\mu f_2 
+ f_{1}^* \partial_\mu \phi
- (\partial_\mu \phi^*) f_2 - (\partial_\mu f_1^*) \phi \; \bigr ]
\nonumber\\
&-& 2 e \bigl [\; A_\mu\; (\phi^* f_2 + f_1^* \phi) - i (\partial_\mu C) 
\phi^* \phi\; \bigr ] \nonumber\\
\bar K_\mu^{(c)} &=& i\; \bigl [ \;\phi^* \partial_\mu 
 \bar f_1 + \bar f^*_2 \partial_\mu
\phi - (\partial_\mu \phi^*) \bar f_1 - (\partial_\mu \bar f^*_2)
 \phi \;\bigr ]
\nonumber\\
&-& 2 e\; \bigl 
[\; A_\mu (\phi^* \bar f_1 + \bar f^*_2 \phi) - i (\partial_\mu \bar C)
(\phi^* \phi) \;\bigr ].
\end{array} \eqno(5.4)
$$
The explicit expression for $L_{\mu 1}^{(c)}$ and
$L_{\mu 2}^{(c)}$, in the above equation,  are
$$
\begin{array}{lcl}
L_{\mu 1}^{(c)} &=& i\; \bigl [\;\phi^* \partial_\mu b + b^* \partial_\mu \phi
+ i\; (f_{1}^* \partial_\mu \bar f_{1} - \bar f_{2}^* \partial_\mu f_2)
\nonumber\\
&-& (\partial_\mu \phi^*) b - (\partial_\mu b^*) \phi + i\; 
\{ \;(\partial_\mu 
\bar f_2^*) f_2 - (\partial_\mu f_1^*) \bar f_1)\;\} \;\bigr ] \nonumber\\
L_{\mu 2}^{(c)} &=& - 2 \;e\;\bigl [\; (A_\mu)\; \bigl (\; \phi^* b + b^* \phi
+ i f_{1}^* \bar f_1 - i \bar f_2^* f_2\;\bigr ) - (\partial_\mu \bar C)
\;(\phi^* f_2 + f_1^* \phi) \nonumber\\
&-& (\partial_\mu C)\; (\phi^* \bar f_1 + \bar f_2^* \phi) \; + \; 
(\partial_\mu B)\; (\phi^* \phi)\; \bigr ].
\end{array} \eqno(5.5)
$$
In sections 3 and 4, we have been able to show that the
nilpotent ($Q_{(a)b}^2 = 0$) (anti-)BRST 
charges $Q_{(a)b}$ that generate the nilpotent ($s_{(a)b}^2 = 0$)
transformations correspond to the translation generators
$(\mbox{ Lim}_{\bar\theta \rightarrow 0} \partial /\partial \theta))
\mbox{ Lim}_{\theta \rightarrow 0} (\partial /\partial \bar \theta)$
along the Grassmannian $(\theta)\bar\theta$-directions of the supermanifold.
This statement is valid for the derivation of the nilpotent
symmetry transformations for the gauge, (anti-)ghost and matter
fields of any given interacting gauge theory in the framework of
the {\it augmented} superfield formalism. We christen our present
superfield formalism, where the horizontality condition
and the invariance of the matter (super)currents on the (super)manifolds
are exploited together, as the {\it augmented} superfield formalism. 
To maintain the sanctity
of this geometrical interpretation for the case of any arbitrary
fields (eg, the composite fields $\tilde J_\mu^{(c)}$), it
is straightforward to re-express the most general expansion (5.3) as
$$
\begin{array}{lcl} 
\tilde J_\mu^{(c)}\; (x, \theta, \bar\theta) 
= J_\mu^{(c)}\; (x) + \theta \; (s_{ab} J_\mu^{(c)} (x))
+ \bar \theta \; (s_b J_\mu^{(c)} (x)) 
+ \theta\; \bar\theta\; (s_{b} s_{ab} J_\mu^{(c)} (x)). 
\end{array} \eqno(5.6)
$$
It can be readily verified
that $s_{(a)b} J_\mu^{(c)} = 0$ where the conserved ordinary
matter current $J_\mu^{(c)} (x) \sim \phi^{*} D_\mu \phi
- \phi D_\mu \phi^{*}$ (cf section 2) and $s_{(a)b}$ are the
off-shell nilpotent (anti-)BRST transformations in (2.6). 
Insertions of these explicit values
(ie $s_b J_\mu^{(c)} = 0, s_{ab} J_\mu^{(c)} = 0$) in (5.6) imply the natural
equality of the super matter current and the ordinary matter current
(ie $\tilde J_\mu^{(c)} (x,\theta,\bar\theta) = J_\mu^{(c)} (x)$) because
all the individual terms on the rhs of (5.6) vanish. The comparison
between (5.6) thus obtained and the general expansion in (5.3) leads to
the following restrictions
$$
\begin{array}{lcl}
s_{ab} J_\mu^{(c)} = \bar K_\mu^{(c)} = 0 \;\qquad \; 
s_b J_\mu^{(c)} = K_\mu^{(c)} = 0 \;\qquad \;
s_b s_{ab} J_\mu^{(c)} = L_\mu^{(c)} = L_{\mu 1}^{(c)} + L_{\mu 2}^{(c)} = 0.
\end{array} \eqno(5.7)
$$
A careful look at the expressions in (5.4) and (5.5)
leads to the following solutions for the restrictions (5.7) in terms
of the (anti-)ghost fields $(\bar C)C$ and the matter fields
\footnote{It is interesting to note that for the condition
$s_b J_\mu^{(c)} = K_\mu^{(c)} = 0$ to be satisfied (cf (5.7)),
it is clear that the {\it odd} looking term $A_\mu (\phi^* f_2 + f_1^* \phi)$
in (5.4)
should be zero on its own. This can be easily achieved if the fermionic
secondary fields $f_2$ and $f_1^*$ are proportional to the basic bosonic
fields $\phi$ and $\phi^*$ respectively. To make the latter pair 
fermionic in nature, a smart guess is $f_2 \sim - C\phi,
f_1^* \sim C \phi^*$. Exactly the same kind of argument is
valid for $s_{ab} J_\mu^{(c)} = \bar K_\mu^{(c)} = 0$ which entails upon
the secondary fields to be:
$ \bar f_1 \sim - \bar C \phi, \bar f_2^* \sim \bar C \phi^*$. 
The rest of the choices in (5.8) follow exactly the similar kind of logical
arguments.}
$$
\begin{array}{lcl}
&&\bar f_1 = - e \bar C \phi\; \qquad f_2 = - e C \phi\; \qquad
\bar f_2^* = + e \bar C \phi^*\; \qquad f_{1}^* = + e C \phi^*
\nonumber\\
&& b^* = i \;e\;\phi^* \bigl [ B - e \bar C C\bigr ]\; \qquad\;\;
 b = - i \;e\; \bigl [ B + e \bar C C\bigr ] \; \phi.
\end{array} \eqno(5.8)
$$
The explicit computation, with the above insertions,
leads to the precise expression for $L_{\mu 1}^{(c)} 
= 2 e (\partial_\mu B) (\phi^* \phi)$ which exactly cancels with the 
computed  value of $L_{\mu 2}^{(c)}$, given by
$L_{\mu 2}^{(c)} = - 2 e (\partial_\mu B) (\phi^* \phi)$. Rest of
the conditions are also very beautifully satisfied
which finally lead to the restriction on the supermanifold as
$\tilde J_{\mu}^{(c)} (x, \theta, \bar\theta) = J_\mu^{(c)} (x)$. 
We wish to re-emphasize that this condition is {\it not} put by hand
from outside. It is the inherent property of the theory itself. In other words,
the off-shell nilpotent symmetries (2.6) for the matter fields are such
that the supercurrent $\tilde J_\mu^{(c)} (x,\theta,\bar\theta)$,
even though expanded along $\hat {\bf 1}, \theta, \bar\theta$ and 
$\theta\bar\theta$-directions of the six dimensional supermanifold,
gets rid of its Grassmannian dependence and reduces to its local
version $J_\mu^{(c)} (x)$ on the 4D manifold.
Ultimately, the super expansion in (5.1), in the light of (5.8), becomes
$$
\begin{array}{lcl} 
 \Phi (x, \theta, \bar\theta) &=& \phi (x)
+ \theta\; (s_{ab} \phi (x)) +  \;\bar \theta \; (s_b \phi (x)) 
+ \theta \;\bar \theta (s_b s_{ab} \phi (x))
\nonumber\\
\bar \Phi^* (x, \theta, \bar\theta) &=&  \phi^* (x)
+ \theta \;(s_{ab} \phi^* (x)) + \bar \theta \; (s_b \phi^* (x)) 
+ \theta \;\bar \theta (s_b s_{ab} \phi^* (x)).
\end{array} \eqno(5.9)
$$
It is clear that the analogue of (4.9) can be written for the
interacting $U(1)$ gauge theory involving the complex scalar fields with the
replacements $\Sigma (x) = \phi (x), \phi^* (x)$ and
$\tilde \Sigma (x, \theta,\bar\theta) = \Phi (x,\theta,\bar\theta),
\Phi^* (x,\theta,\bar\theta)$. In a similar fashion, the analogue of
(4.10) is also valid for the system of complex scalar fields in
interaction with the $U(1)$ gauge field $A_\mu$.\\

\noindent
{\bf 6 Conclusions}\\

\noindent
In our present investigation, we have addressed the long-standing problem of
the derivation of the off-shell nilpotent (anti-)BRST symmetry
transformations for the matter fields, present in the interacting 
Abelian $U(1)$ gauge theories, in the framework
of augmented superfield formalism. The
field theoretical examples that we have chosen are (i) the Dirac fields
in interaction with the $U(1)$ gauge field $A_\mu$, and (ii) the interacting
Abelian 1-form gauge theory involving the complex scalar fields as the matter 
fields. As it turned out,
for both the above cases of the interacting field theories, it is the
requirement of the invariance of the conserved (super)currents, defined
on the (super)manifolds, that is responsible for the derivation of
the off-shell nilpotent (anti-)BRST transformations on the
matter fields. There is a clear-cut distinction, however, between the
mechanism of derivation of the above symmetries for the cases of 
(i) the Dirac fields, and (ii) the complex scalar fields. For the case
of the interacting Abelian gauge theory with the Dirac fields, the 
horizontality condition (which is responsible for the derivation of the 
nilpotent symmetries for the gauge- and (anti-)ghost fields (cf section 3)),
{\it does not} play any significant role in the derivation of the
corresponding nilpotent (anti-)BRST
symmetry transformations for the matter fields 
(see, eg, section 4). This is due to the fact that the matter supercurrent
$\tilde J_\mu^{(d)} (x,\theta,\bar\theta) 
= \bar\Psi (x,\theta,\bar\theta) \gamma_\mu 
\Psi (x,\theta,\bar\theta)$ does not contain any superfields corresponding
to the basic fields $A_\mu, C, \bar C$.
On the contrary, the horizontality condition does
play a very crucial and decisive role in the derivation of the
nilpotent (anti-)BRST 
symmetry transformations for the complex scalar fields. This is primarily
because of the fact that the conserved supercurrent 
$\tilde J_\mu^{(c)} (x,\theta,\bar\theta)$ (cf (5.2)) contains the superfield 
$B_\mu (x,\theta,\bar\theta)$ corresponding to the gauge field $A_\mu (x)$.
While computing the super expansion for the 
$\tilde J_\mu^{(c)} (x,\theta,\bar\theta)$ along 
$\hat {\bf 1}, \theta, \bar\theta$
and $\theta\bar\theta$-directions of the supermanifold, we do require
the expansion in (3.6) for the superfield
$B_\mu (x,\theta,\bar\theta)$ which is derived after the restriction 
(i.e. $\tilde F = F$) due to the horizontality condition is imposed
on the (super) curvature 2-forms.

On the face value, it appears very surprising that the off-shell nilpotent
transformations for the gauge- and (anti-)ghost fields
derived from the horizontality condition are consistent with and complementary
to the nilpotent transformations for the matter fields
derived from the requirement of 
the invariance of the conserved matter (super)currents of the theory. However,
there is an explanation for this mutual consistency 
and complementarity between the two. In fact,
the nilpotent (anti-)BRST transformations for the gauge fields
(that involve the (anti-)ghost
fields) are encoded in the curvature 2-form $F = d A$ for the Abelian
$U(1)$ gauge theory as it remains invariant under 
the (anti-)BRST transformations $A \rightarrow A + d \bar C,
A \rightarrow A + d C$. This is why when we demand $\tilde F = F$
on the six $(4 + 2)$-dimensional supermanifold, we obtain the transformations
on the gauge- and (anti-)ghost fields. To express the same thing in 
the physical language, we just demand that the 
classical physical (i.e. BRST invariant) fields ${\bf E}$ and ${\bf B}$, in the
superfield formulation, {\it should not}
get any contribution from the Grassmann variables. The next physically
important object in the interacting $U(1)$ gauge 
theory is the matter conserved current which plays a significant role in
the interaction term $J_\mu^{(c, d)} A^\mu$
(where the matter conserved current couples to the gauge field). 
The conserved matter current is derived due to the global gauge
invariance in the theory (Noether theorem). However, the
interaction term owes its origin to the requirement of the
{\it local} gauge invariance (gauge principle [41]). Thus, the outcome 
from the requirement of the {\it invariance} of the matter (super)currents 
on the (super)manifolds is 
mutually consistent with and complementary to the local gauge (i.e. BRST)
invariance  of the curvature 2-form $F = d A$ as the principle of {\it local}
gauge invariance is the common and connecting thread that
runs through both of the above requirements.

In the present paper, we have concentrated only on the local, covariant,
continuous and off-shell nilpotent (anti-)BRST symmetry transformations
for the matter fields. However, for the Dirac fields in interaction with
the $U(1)$ gauge field $A_\mu$, it is already known that there exists
a set of non-local, non-covariant, continuous and nilpotent (anti-)co-BRST
symmetry transformations (see, eg, [34] for detailed references).
Such kind of symmetries for the Dirac fields have also been found for the
non-Abelian gauge theory where there is a coupling between the $SU(N)$
gauge field and the matter (Dirac) conserved current [42]. It would be
interesting endeavour to extend our present work to include these
non-local and non-covariant symmetry transformations. Furthermore, our
present investigation can be generalized readily to the
{\it interacting } non-Abelian gauge theory where the local, covariant, 
continuous and off-shell nilpotent (anti-)BRST transformations do exist
for the non-Abelian gauge field, the (anti-)ghost fields and the Dirac fields. 
The derivation of the on-shell version of the above symmetries 
for the interacting (non-)Abelian gauge theories with matter fields
is another direction that can be pursued in 
the future. These are some of the issues that are under investigation and
our results will be reported elsewhere [43]. \\

\noindent
{\bf Acknowledgements}\\

\noindent
The critical and constructive comments by the referees are gratefully
acknowledged.

\end{document}